\documentclass[10pt,twocolumn,twoside,letterpaper]{IEEEtran}

\usepackage{geometry}
\makeatletter
\def\ps@IEEEtitlepagestyle{
  \def\@oddfoot{\mycopyrightnotice}
  \def\@evenfoot{}
}
\def\mycopyrightnotice{
  {\footnotesize
  \begin{minipage}{\textwidth}
  \centering
  978-1-6654-8872-3/22/\$31.00 \copyright2022 IEEE
  \end{minipage}
  }
}

\usepackage{amsmath,amssymb,amsfonts}
\usepackage{graphicx}
\usepackage{textcomp}
\usepackage{xcolor}

\usepackage{hyperref}
\usepackage{graphicx}

\usepackage{times}
\usepackage{xspace}
\usepackage[colorinlistoftodos]{todonotes} 
\usepackage{bm} 
\usepackage{subcaption}

\usepackage{epstopdf}
\AppendGraphicsExtensions{.tiff}
\graphicspath{{fig/}} 

\usepackage{epsfig}
\usepackage{tikz}
\usepackage{algpseudocode}
\usepackage[linesnumbered, ruled,vlined]{algorithm2e}
\usepackage{mathrsfs}

\usepackage[numbers, sort&compress]{natbib}   
\usepackage{hyperref}

\def\QED{~\rule[-1pt]{5pt}{5pt}\par\medskip}

\DeclareMathAlphabet{\altmathcal}{OMS}{cmsy}{m}{n} 

\newcommand{\st}{\operatorname{s.t.}}

\newcommand{\xmath}[1] {\ensuremath{#1}\xspace}
\newcommand{\blmath}[1] {\xmath{\bm{#1}}}
\newcommand{\A}{\blmath{A}}

\newcommand{\D}{\blmath{D}}

\newcommand{\x}{\blmath{x}}
\newcommand{\y}{\blmath{y}}

\newcommand{\va}{\blmath{a}}
\newcommand{\vb}{\blmath{b}}
\newcommand{\vc}{\blmath{c}}
\newcommand{\vd}{\blmath{d}}

\newcommand{\vw}{\blmath{w}}
\newcommand{\vs}{\blmath{s}}
\newcommand{\veps}{\blmath{\epsilon}}

\newcommand{\1}{\blmath{1}}

\newcommand{\conv}{\circledast}

\newcommand{\norm}[2]{\left\| #1 \right\|_{#2}}
\newcommand{\abs}[1]{\left| #1 \right|}
\newcommand{\innerprod}[2]{\left\langle #1,  #2 \right\rangle}

\renewcommand{\baselinestretch}{1.020}
\renewcommand{\mathbf}{\boldsymbol}

\usepackage{url}

\begin{document}
\title{Fusing Sparsity with Deep Learning for  Rotating Scatter Mask Gamma Imaging
}
%
%
%

\author{Yilun Zhu, \IEEEmembership{Student Member, IEEE}, Clayton D. Scott, \IEEEmembership{Senior Member, IEEE}, 
Darren E. Holland, \\ George V. Landon, \IEEEmembership{Member, IEEE}, Aaron P. Fjeldsted, \IEEEmembership{Student Member, IEEE}, Azaree T. Lintereur, \IEEEmembership{Member, IEEE}
     
\thanks{This work was supported in part by the Department of Defense, Defense Threat Reduction Agency under award HDTRA1-20-2-0002.}

\thanks{Y. Zhu and C. D. Scott are with the Department of Electrical Engineering and Computer Science, University of Michigan (e-mail: \{allanzhu, clayscot\}@umich.edu).}

\thanks{D. E. Holland is with the Department of Engineering Physics, Air Force Institute of Technology (e-mail: Darren.Holland@afit.edu).}
\thanks{G. V. Landon is with the School of Engineering and Computer Science, Cedarville University (e-mail: georgelandon@cedarville.edu).}

\thanks{A. P. Fjeldsted, A. T. Lintereur are with the Ken and Mary Alice Lindquist Department of Nuclear Engineering, Penn State University (e-mail: \{apf5504, atl21\}@psu.edu).}
}

\maketitle

\pagenumbering{gobble}

\begin{abstract}
    Many nuclear safety applications need fast, portable, and accurate imagers to better locate radiation sources. The Rotating Scatter Mask (RSM) system is an emerging device with the potential to meet these needs.  The main challenge is the under-determined nature of the data acquisition process: the dimension of the measured signal is far less than the dimension of the image to be reconstructed. To address this challenge, this work aims to fuse model-based sparsity-promoting regularization and a data-driven deep neural network denoising image prior to perform image reconstruction. An efficient algorithm is developed and produces superior reconstructions relative to current approaches.
\end{abstract}


\section{Introduction}
\IEEEPARstart{L}{ocalizing} radioactive material is particularly important for nuclear safety applications, such as environmental monitoring and searching for orphan sources. There is growing need for portable imaging systems that give accurate results in real-time. Coded aperture systems~\cite{vetter2006gamma} offer such capabilities, but have a limited field-of-view (FOV). Recently, a novel time-encoded gamma-ray detection system called the Rotating Scatter Mask (RSM)~\cite{fitzgerald2015rotating} with a near $4\pi$ FOV was developed.

For the emerging RSM system, novel and suitable image reconstruction methods are needed. One of the biggest challenges of RSM imaging is the limited number of measurements: the goal is to recover an image of size $m \times n$ from $n$ measurements. Previously, Olesen et al. \cite{olesen2020maximum} proposed an optimization-based method by using maximum-likelihood expectation–maximization (ML-EM) with median root prior \cite{alenius1997bayesian}. Later, they developed an end-to-end learning-based method \cite{olesen2021regenerative} that exhibited improved performance. 

This paper leverages both analytical sparsity-promoting $\ell_1$ regularization and data-driven neural network-based smoothness prior. Combining these two regularizations is valid due to the structure of images of interest: sparse radiation sources that are located in containers with regular geometries. 

\section{Rotating Scatter Mask Imaging}

\begin{figure}[h]
    \centering
    \includegraphics[width=0.6\linewidth]{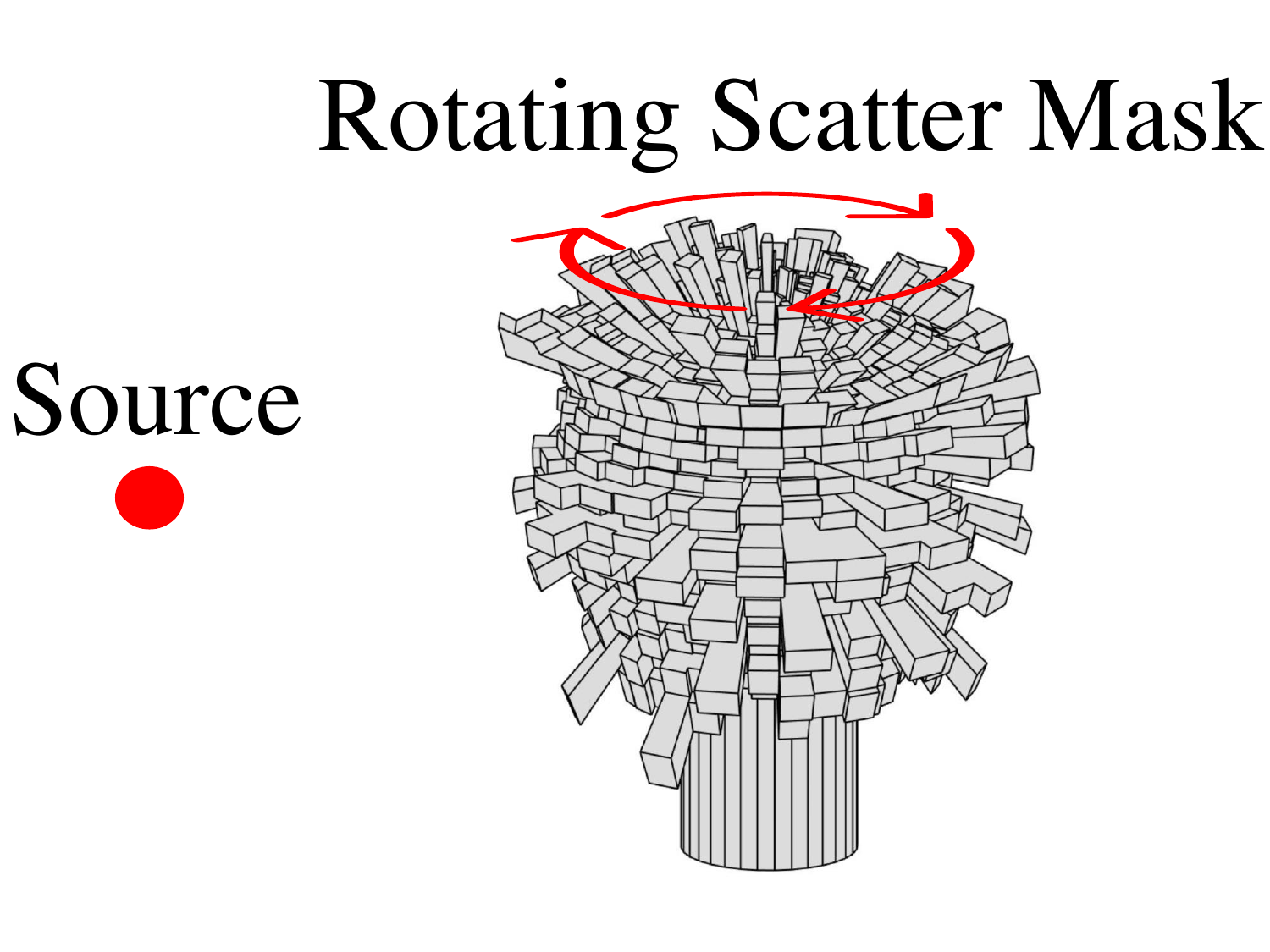}
    \caption{Rotating Scatter Mask system, with an interior detector that measures a time/angle varying signal. Figure used with permission \cite{olesen2020low}. }
    \label{fig:RSM}
\end{figure}

Fig. \ref{fig:RSM} shows the schematic of an RSM system. A single scintillating detector is covered by a mask made of homogeneous poly(methyl methacrylate) (PMMA)  \cite{olesen2020maximum}. As the mask spins, the recorded measurement is a time-varying noisy signal $\y = \left[y_1 \ldots y_n\right]'$ called the detector response curve (DRC), viewed as a column vector. For simplicity, assume the mask rotates one cycle and $n$ denotes the number of mask elements along the horizontal direction. At time index $i$, $y_i$ represents the total number of gamma rays detected during the $i$-th time interval. 

The detector response matrix (DRM) is expressed via an $m \times n$ matrix $\D$, where $m$ is the total number of mask elements along the vertical direction. Each element $\D_{ij}$ represents the expected detector response due to a source located at the center of the $(i,j)$-th discretized mask voxel ~\cite{holland2018rotating}. The image $\A$ to be reconstructed is also $m \times n$.  Mathematically, the measured DRC is produced by convolving an image with the DRM along the horizontal direction.  The ideal measured DRC, denoted $\vs$, can thus be expressed as a sum of 1D convolutions
\begin{equation}
    \vs = \sum_{i=1}^m \D_{i,:} \conv \A_{i,:} =: \Phi \va,
\end{equation}
where $\D_{i,:}$ and $\A_{i,:}$ denote the $i$-th row of matrices $\D$ and $\A$, respectively, $\Phi$ is the concatenation of circulant matrices generated from $\D_{i,:}$, and $\va$ is the vectorized image. The actual measured signal $\y$ is a noisy version of the ideal signal $\vs$.

The challenge of this image reconstruction task stems from the limited dimension of the measured signal compared to image size. Romberg~\cite{romberg2009multiple} has analyzed a similar problem and found a sufficient condition of sparse image recovery to be $n > O \left( \log^4{mn} \right)$. However, this condition is violated for RSM imaging tasks when $m=75$ and $n=180$.
This violation indicates using sparsity alone is not enough and more advanced image reconstruction algorithms are needed.

\section{Alternating direction method of multipliers (ADMM) $\&$ Plug-and-Play (PnP)}
Image $\va$ is estimated from measurement $\y$ by means of regularization-based image reconstruction, which seeks the image that best explains the observations, while also conforming to prior expectations. To be specific, the goal is to solve an optimization problem 
\begin{equation}
    \min_{\va} f(\va) + h(\va),
    \label{eq:uncons}
\end{equation}
with $f(\va)$ aiming to fit data and $h(\va)$ encoding image prior. This paper focuses on using Alternating direction method of multipliers (ADMM) \cite{eckstein1992douglas, boyd2011distributed} to solve the above objective function.
In this section, we provide a general overview of ADMM, and in the next section, we apply this technique for our specific model.

\subsection{Standard ADMM}
The idea of ADMM is to convert \eqref{eq:uncons} into a constrained form by variable splitting
\begin{equation}
    \min_{\va,\vb} f(\va) + h(\vb), \quad \st \quad \va=\vb,
\end{equation}
and consider its augmented Lagrangian form 
\begin{equation}
    L_\rho(\va,\vb,\vw) = f(\va) + h(\vb) + \innerprod{\va-\vb}{\vw} + \frac{\rho}{2} ||\vb-\va||_2^2,
    \label{eq:aug_lag}
\end{equation}
where $\vw$ is the Lagrange multiplier and $\rho > 0$.
Then, minimize \eqref{eq:aug_lag} by solving a sequence of sub-problems 
\begin{align}
    \va^{(k+1)} & \leftarrow \arg \min_{\va} L_\rho \left(\va,\vb^{(k)},\vw^{(k)}\right),  \\
    \vb^{(k+1)} & \leftarrow \arg \min_{\vb} L_\rho \left(\va^{(k+1)},\vb,\vw^{(k)}\right), \label{eq:b_update} \\ 
    \vw^{(k+1)} & \leftarrow \vw ^k+ \rho  \left(\va^{(k+1)}-\vb^{(k+1)}\right),
\end{align}
where the update of each sub-problem often has a closed-form solution for various choices of $f(\cdot)$ and $h(\cdot)$.

\subsection{Plug-and-Play ADMM}
The modular structure of ADMM updates allows us to incorporate powerful data-driven methods into the optimization scheme. To be specific, \eqref{eq:b_update} can be viewed as an image denoising problem as shown in \eqref{eq:denoise} and the image prior $h(\cdot)$ is encoded via the proximal operator $ H_\sigma(\cdot)$ associated with it in \eqref{eq:prox}, where $\sigma = \sqrt{ 1/\rho }$.
\begin{align}
    \vb^{(k+1)} & \leftarrow  \arg \min_{\vb}  \frac{1}{2\sigma^2} \norm{\vb-(\va^{(k+1)}+\vw^{(k)})}{2}^2 + h(\vb), \label{eq:denoise} \\ 
    & = H_\sigma\left(\va^{(k+1)}+\vw^{(k)}\right) \label{eq:prox}.
\end{align}
For example, if $h(\vb) = \norm{\vb}{1}$, then $H_\sigma (\cdot)$ is the soft-thresholding operator. Building upon this, Venkatakrishnan et al. \cite{venkatakrishnan2013plug} proposed the Plug-and-Play (PnP) ADMM framework by plugging advanced image denoising methods into $H_\sigma (\cdot)$, which do not necessarily correspond to an analytical image prior. This approach produces state-of-the art results in many imaging tasks \cite{rond2016poisson, chan2016plug, ahmad2020plug, sun2021plug}.

\section{Proposed Method}
We propose combining sparsity with deep learning for RSM imaging. The data fitting is assessed by calculating the squared error. The \emph{a priori} belief is reflected by a sparsity-promoting $\ell_1$ penalty with a ``deep denoiser prior'' \cite{drunet} represented by a convolutional neural network (CNN)
\begin{equation}
    \min_{\va} \frac{1}{2} \norm{\y - \Phi \va}{2}^2 + \lambda \norm{\va}{1} + \gamma \norm{\va}{\text{CNN}},
\end{equation}
where $\lambda, \gamma$ are hyperparameters, and $\norm{\cdot}{\text{CNN}}$ is a pseudo-prior represented by a CNN \cite{drunet}. Previous works using deep denoiser priors have not addressed sparse images like those arising in nuclear security applications, thus this approach is novel. The above optimization problem is solved by using the ADMM by first doing variable splitting 
\begin{equation}
    \min_{\va, \vb, \vc} \frac{1}{2} \norm{\y - \Phi \va}{2}^2 + \lambda \norm{\vb}{1} + \gamma \norm{\vc}{\text{CNN}} \quad \st \quad \va = \vb, \va = \vc,
\end{equation}
and then writing out the augmented Lagrangian
\begin{align}
    L_{\rho_1, \rho_2}(\va,\vb, \vc, \vw_1, \vw_2) & = \frac{1}{2} \norm{\y - \Phi \va}{2}^2  + \lambda \norm{\vb}{1} + \gamma \norm{\vc}{\text{CNN}} \nonumber \\
        & + \innerprod{\vb-\va}{\vw_1} + \frac{\rho_1}{2} ||\vb-\va||_2^2 \nonumber  \\
        & + \innerprod{\vc-\va}{\vw_2} + \frac{\rho_2}{2} ||\vc-\va||_2^2.
\end{align}


Our alternating minimization algorithm successively updates according to the data-fitting, sparse promoting and CNN denoising terms, respectively
\begin{align}
    \va^{(k+1)} & \leftarrow \left( \Phi'\Phi + \rho_1 I  + \rho_2 I \right)^{-1} \cdot  \nonumber \\ 
        & \left[ \Phi' \y + \rho_1 \left( \vb^{(k)} + \frac{1}{\rho_1} \vw_1^{(k)} \right) + \rho_2 \left( \vc^{(k)} + \frac{1}{\rho_2} \vw_2^{(k)} \right) \right], \label{eq:inv} \\ 
    \vb^{(k+1)} & \leftarrow S_{\lambda/\rho_1} \left[ \va^{(k+1)} - \frac{1}{\rho_1} \vw_1^{(k)} \right], \\
    \vc^{(k+1)} & \leftarrow \text{Denoiser}_{\gamma/ \rho_2} \left[ \va^{(k+1)} - \frac{1}{\rho_{2}} \vw_2^{(k)}  \right],
\end{align}
where $S_{\lambda/\rho_1} [\cdot]$ denotes the soft-thresholding operator with threshold $\lambda/\rho_1$ and $\text{Denoiser}_{\gamma/ \rho_2}(\cdot)$ is a pre-trained CNN denosier proposed in \cite{drunet}, which takes a noisy image and noise variance $\gamma/ \rho_2$ as input.

Equation~\eqref{eq:inv} requires heavy computation of $\mathcal{O}(m^2 n^2 \log mn)$ because it involves inverting a matrix of size $mn \times mn$. Fortunately, by leveraging the convolutional structure in $\Phi$, the algorithm can be substantially accelerated to $\mathcal{O}(m n \log n)$. Through the use of the matrix inversion lemma and discrete Fourier Transform, \eqref{eq:inv} becomes
\begin{align}
    \va^{(k+1)} \leftarrow  \frac{1}{\rho} \x^{(k)} - \frac{1}{\rho^2}  \Phi' \mathcal{F}^{-1}  \left\{ \frac{\mathcal{F} \left( \Phi \x^{(k)} \right) }{\1 + \frac{1}{\rho} \sum\limits_{i=1}^m \abs{ \mathcal{F}(\vd_i) }^2} \right\},
\end{align}
where $\mathcal{F}(\cdot)$ denotes the Fourier transform, $\vd_i$ is the  $i$-th row of the DRM, $\x^{(k)} := \Phi' \y + \rho_1 \left( \vb^{(k)} + \frac{1}{\rho_1} \vw_1^{(k)} \right) + \rho_2 \left( \vc^{(k)} + \frac{1}{\rho_2} \vw_2^{(k)} \right)$, $\rho = \rho_1 + \rho_2$ and the operations inside inverse Fourier transform are performed element-wisely. The overall algorithm is shown in Algorithm \ref{alg:fsdl}.

\begin{algorithm}[h]
    \SetAlgoLined      
    \KwIn{ measurment $\y$, system matrix $\Phi$, Deep denoiser model, number of iterations $K$, hyperparameter $\lambda, \gamma, \rho_1, \rho_2 > 0$. }
    
    Initialize $\va^{(0)}, \vb^{(0)}, \vc^{(0)}, \vw_1^{(0)}, \vw_2^{(0)} $ as zero. \\
    \For{$k = 0$ \KwTo $K-1$}{
        $ \va^{(k+1)} \leftarrow  \frac{1}{\rho} \x^{(k)} - \frac{1}{\rho^2}  \Phi' \mathcal{F}^{-1}  \left\{ \frac{\mathcal{F} \left( \Phi \x^{(k)} \right) }{\1 + \frac{1}{\rho} \sum_{i=1}^m \abs{ \mathcal{F}(\vd_i) }^2} \right\}, \newline \text{ where } \x^{(k)} := \Phi' \y + \rho_1 \left( \vb^{(k)} + \frac{1}{\rho_1} \vw_1^{(k)} \right) + \rho_2 \left( \vc^{(k)} + \frac{1}{\rho_2} \vw_2^{(k)} \right), \ \rho = \rho_1 + \rho_2 $  \;
        $ \vb^{(k+1)} \leftarrow S_{\lambda/\rho_1} \left[ \va^{(k+1)} - \frac{1}{\rho_1} \vw_1^{(k)} \right] \newline \vw_1^{(k+1)} \leftarrow \vw_1^{(k)} + \rho_1 \left( \vb^{(k+1)} - \va^{(k+1)} \right) $ \;
        $ \vc^{(k+1)} \leftarrow \text{Denoiser}_{\gamma/ \rho_2} \left[ \va^{(k+1)} - \frac{1}{\rho_{2}} \vw_2^{(k)}  \right] \newline 
        \vw_2^{(k+1)} \leftarrow \vw_2^{(k)} + \rho_2 \left( \vc^{(k+1)} - \va^{(k+1)} \right) $ \;
    }

    \KwOut{Reconstructed image $ \widehat{\va} = \va^{(K)} $.}

    \caption{RSM image reconstruction by fusing sparsity with deep learning. $S[\cdot]$ denotes soft-thresholding operator. $\vw_i$ refers to Lagrange multipliers that arise within the ADMM framework, one for each penalty term.}
    \label{alg:fsdl}
\end{algorithm}

\section{Experiments}
In this section, results of the proposed algorithm are presented for synthetically generated noisy data.
\subsection{Experimental Setup}
The 1'' cylindrical CsI detector's response (i.e., elements of DRM) within the full-energy peak was simulated using 662~keV photons located 86.36~cm from the detector center in Monte Carlo N-Particle (MCNP) 6.2.0 for every mask voxel with enough particles to produce a relative error below 3\% for each position.  

The algorithm was tested on a suite of 20 sparse, but distributed images that contain three categories of shapes (disc, ring, and square). These shapes mimic different cross-sections of nuclear waste containers and contaminated storage drums. Among 20 testing images, there were 6 discs, 6 rings and 8 squares, each located in different angular positions (which leads to the distortion of regular shapes caused by viewing from spherical coordinate system) and having different sizes.

For a given true image $\va$, $\y = \Phi \va + \veps$ was simulated, where $\epsilon_i \sim {\cal N}(0,\sigma^2)$, such that $\y$ has an average intensity of $10,000$ counts, and where $\sigma^2 = 10,000$. Since the average intensity is large, the Gaussian noise model is a good approximation of Poisson measurements.

The quality of the reconstructed result on a specific image was evaluated via normalized root mean square error (NRMSE)
\begin{equation*}
    \text{NRMSE} = \frac{\sqrt{\frac{1}{mn} \sum_i \abs{\widehat{\va}[i] - \va[i]}^2}}{\sqrt{\frac{1}{mn} \sum_i \abs{\va[i]}^2}} = \frac{\norm{\widehat{\va} - \va}{2}}{\norm{\va}{2}},
\end{equation*}
where $\va$ is the ground truth image and $\widehat{\va}$ is the estimated image. 

\subsection{Choice of Parameters}
We tuned the hyperparameter of the proposed algorithm on a validation dataset, leading to 
\begin{equation*}
    \lambda = 0.36, \quad \gamma = 0.23, \quad K = 300.
\end{equation*} 
As for the choice of $\rho_1, \rho_2$, we set them equal $\rho^{(k)} := \rho_1^{(k)} = \rho_2^{(k)}$ and adopt the common strategy of gradually increasing the value as a function of iteration index $k$ \cite{drunet}, resulting in $\rho^{(1)} < \cdots < \rho^{(k)} < \cdots < \rho^{(K)}$.

\subsection{Reconstruction Results}
The proposed method ($\ell_1-$CNN) was compared with MLEM-MRP \cite{olesen2020maximum} and $\ell_1$ reconstruction \cite{candes2006stable}. MLEM-MRP was implemented according to \cite{olesen2020maximum} and used the suggested hyperparameters. For $\ell_1$ reconstruction, an iterative solver was implemented via ADMM.

 Table \ref{table:nrmse} shows that $\ell_1-$CNN achieves the best results for all sources of interest. On average, our method improved more than $30 \%$ in terms of NRMSE. Using only the denoising CNN \cite{drunet} without $\ell_1$ regularization did not lead to meaningful results since the reconstructed images were not sparse (therefore not reported in the table).

\begin{table}[h!]
    \centering
    \normalsize
    \begin{tabular}{ c | c c c }
         & MLEM-MRP & $\ell_1$ & $\ell_1-$CNN  \\ \hline
    Disc & 0.703 & 1.41 & \textbf{0.340} \\  
    Ring & 0.894 & 1.15 & \textbf{0.780} \\ 
    Square & 0.787 & 1.35 & \textbf{0.545} \\ \hline   
    Average & 0.794 & 1.31 & \textbf{0.553}
    \end{tabular}
    \caption{Average NRMSE of the three methods evaluated on testing set data. Lower values correspond to improved performance.}
    \label{table:nrmse}
\end{table}

\begin{figure*}[htb]
    \centering

    \begin{subfigure}[b]{\textwidth}
        \includegraphics[width=1\linewidth]{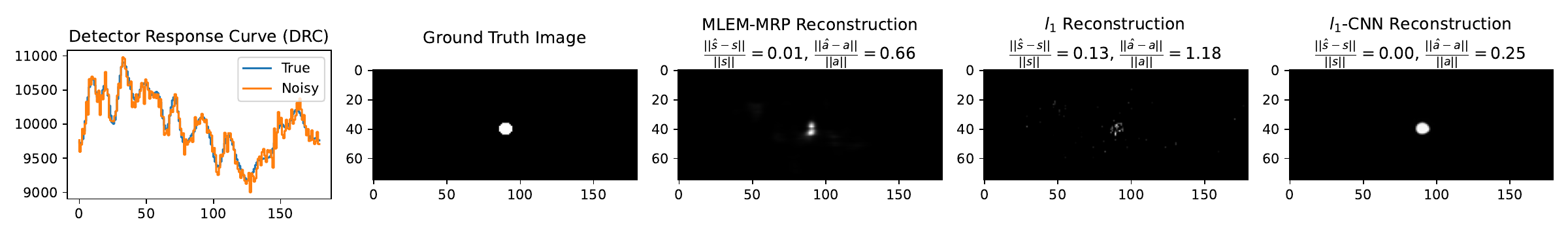}
    \end{subfigure}
    \\[-2ex]
    
    \begin{subfigure}[b]{\textwidth}
        \includegraphics[width=1\linewidth]{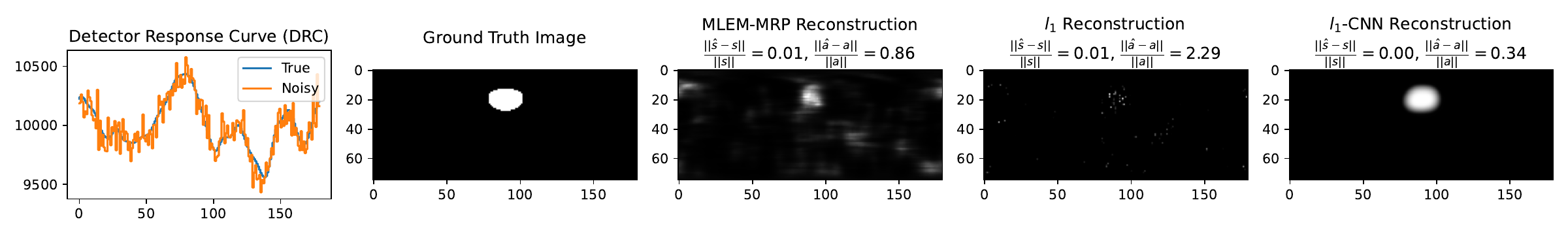}
    \end{subfigure}
    \\[-2ex]

    \begin{subfigure}[b]{\textwidth}
        \includegraphics[width=1\linewidth]{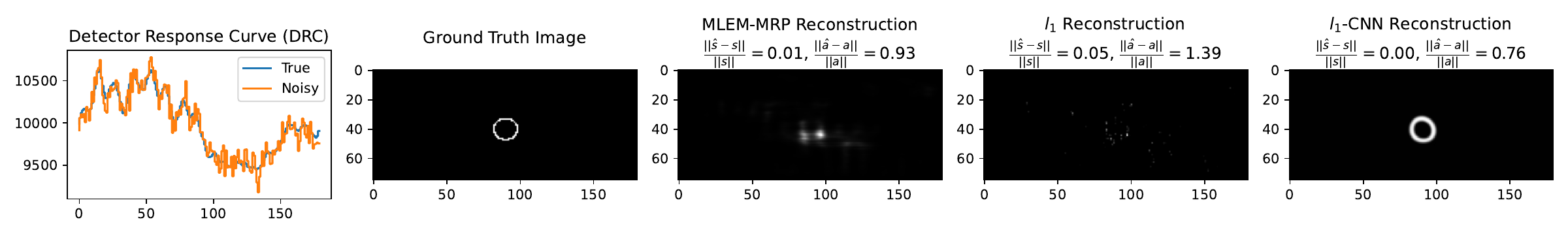}
    \end{subfigure}
    \\[-2ex]

    \begin{subfigure}[b]{\textwidth}
        \includegraphics[width=1\linewidth]{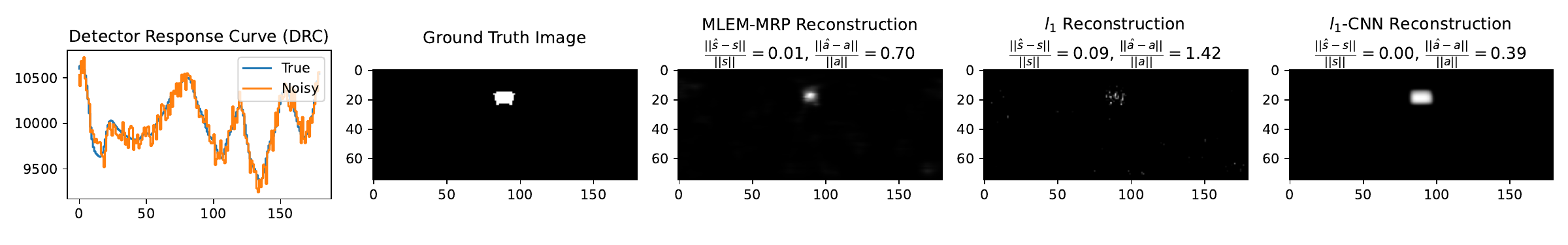}
    \end{subfigure}
    \\[-2ex]

    \begin{subfigure}[b]{\textwidth}
        \includegraphics[width=1\linewidth]{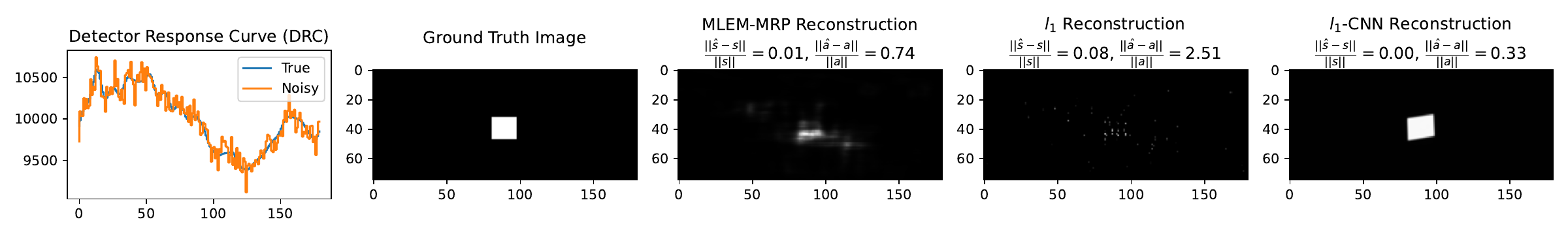}
    \end{subfigure}

    \caption{Example image reconstructions for MLEM-MRP, $\ell_1$ and $\ell_1-$CNN. Left to right: true and received signals, true image and the reconstructed images by the three methods. The NRMSE of each reconstructed signal and image shows the proposed approach achieves the best results quantitatively.}
    \label{fig:lasso_vs_ours}
\end{figure*}

Fig. \ref{fig:lasso_vs_ours} shows 5 representative reconstruction results. Our proposed method ($\ell_1-$CNN) performed best both visually and quantitatively (in terms of NRMSE) in all cases. For the reconstruction of square sources, although the NRMSE were small, visually the artifact was obvious. This left room for further improvement, for example, leveraging deep unrolling \cite{monga2021algorithm} for the iterative algorithm and then training the corresponding architecture end-to-end.

\section{Conclusion}
This paper develops a novel gamma source image reconstruction algorithm for the RSM system. The ADMM algorithm was leveraged to fuse a model-based analytical image prior with the latest data-driven approaches. The results show that the method greatly improves the reconstruction NRMSE over $30 \%$ compared to baseline approaches. Furthermore, the proposed framework is formulated in a general way that has potential to enhance the performance of other imaging systems.

\section*{Acknowledgment}

We thank Maj James E. Bevins for useful discussions.



%





\footnotesize{
  \bibliography{IEEEabrv, refs}
}




\end{document}